\input phyzzx
\hsize=417pt %\vsize=600pt \baselineskip=20pt \maxdepth=0.2pt
\sequentialequations
\Pubnum={ EDO-EP-6}
\date={ June 1996}
\titlepage
\vskip 32pt
\title{ Black hole entropy from a single long string}
\author{Ichiro Oda\footnote \dag{E-mail address: 
sjk13904@mgw.shijokyo.or.jp}}
\vskip 16pt
\address{ Edogawa University,                                
          474 Komaki, Nagareyama City,                        
          Chiba 270-01, JAPAN     }                          
          
%
%              The titlepage ends at this place.
%
%======================================================================%
%
\abstract{In this article we derive the Bekenstein-Hawking formula of 
black hole entropy from a single long string. We consider a open string 
in the Rindler metric which can be obtained in the large mass limit 
from the Schwarzschild black hole metric. By solving the field 
equations we find a nontrivial solution with the exact value 
of the Hawking temperature.  We see that this solution gives us the 
Bekenstein-Hawking formula of black hole entropy to leading order of 
approximation. This string has effectively a rescaled string tension 
having a relation with a redshift factor. It is also pointed out 
that this formalism is extensible to other black holes in an arbitrary
spacetime dimension. The present work might lead us to a surprising 
idea that the recent picture which the black hole entropy arises from 
D-brane excitations has the root of the picture that the black hole 
entropy is stocked in a single long string with a rescaled string tension.
\vskip 16pt
PACS number(s): 11.25.-w, 04.70.Dy
 }
\endpage
% 
%=======================================================================%
%
%        Macros
%

\def\sp(#1){\noalign{\vskip #1pt}}

%

%
%=========================================================================%
%
%   This part is the meat of body.
%
\topskip 30pt
\par
The Bekenstein-Hawking formula of the black hole entropy, 
$S={1 \over 4} {k c^3 \over G \hbar} A_H$ [1,2] contains
the four fundamental constants of physics, which are the Boltzman constant
$k$, the Newton one $G$, the Planck one $\hbar$ and the light velocity $c$ 
so that it suggests a deep triangle relation among thermodynamics, 
general relativity and quantum mechanics. Moreover, this formula relates 
the entropy of a black hole to the area $A_H$ of a event horizon, 
therefore also 
implies a connection to geometry. Thus, although the above formula was 
originally derived in terms of the semi-classical approach, 
it is expected that its validity might be taken over apart from
quantum corrections even when we have a quantum theory of gravitation 
in the future.

However, for more than twenty years the underlying physical basis by which 
${1 \over 4} {k c^3 \over G \hbar} A_H$ arises as the black hole 
entropy remained unclear. It is tempted to regard this black 
entropy as the logarithm of the number of microscopic states compatible 
with the observed macroscopic state from the viewpoint of the ordinary 
statistical physics. Then, a crux of an understanding is what those microscopic
states are. 

Recently this situation has been completely changed by a remarkable work by 
Strominger and Vafa [3]. They showed through an application of D-brane 
method [4] that the Bekenstein-Hawking entropy of an extremal 
five-dimensional black hole precisely equals the number of BPS saturated 
states in string theory with the given charges. Since then a lot of 
related works have appeared [5].

However, it seems to be unsatisfying at least for the present author that 
there is no direct way from string theory of counting the string states 
to match the Bekenstein-Hawking entropy since string theory is believed 
to be a unified theory including a theory of quantum gravity. In 
other words, there should be a physical principle 
from string theory which explains why the number of 
string states is equal to that of statiscal states associated with the 
Bekenstein-Hawking entropy formula in various black hole models 
discovered recently [5]. In this respect, there has more recently 
appeared one interesting idea that black hole entropy might be carried by 
a single long string with a definite central charge and a rescaled string 
tension [6-8].  Motivated by these works, in this paper we would 
like to give an alternate formalism supporting this idea.  

Although Hawking initially showed that black holes evaporate due to 
quantum particle creation and behave as if they had an effective 
temperature given by ${1 \over 8 \pi M}$, with $M$ the mass of the black 
hole, soon after that Unruh showed that a similar thermal radiation can 
be observed in the Rindler spacetime [9]. This is because the Rindler metric 
describes well the essential features of the event horizon of the black 
hole. In fact it is known that in the large mass limit the Rindler     
metric can be obtained from spherically symmetric black hole metric in the 
vicinity of the outer horizon by the coordinate transformations and 
conformal scalings. 

Following these observations, our aim below will be to consider a single 
long string in the Rindler spacetime and to argue that the approach 
developed in [10,11] in the different context does also apply for the 
present problem and leads to qualitative and universal explanation of 
the black hole entropy.

Since we would like to consider the entropy carried by a single long 
string, we start by a bosonic string action in a curved four dimensional
spacetime given by
%%%%%%%%%%%%%%%%%%%%%%%%%%%%Equation%%%%%%%%%%%%%%%%%%%%%%%%%%%%%%%%%%%
$$ \eqalign{ \sp(2.0)
I &= -{T \over 2} \int d^2 \sigma \sqrt{h}
        h^{\alpha\beta} \partial_{\alpha}X^{\mu} \partial_{\beta}
        X^{\nu} g_{\mu\nu}(X),
\cr
\sp(3.0)} \eqno(1)$$
%-----------------------------------------------------------------------
where $T$ is a string tension having dimensions of mass squared, 
$h_{\alpha\beta}(\tau,\sigma)$ denotes the two dimensional world
sheet metric having a Euclidean signature, and $h =det h_{\alpha
\beta}$. $X^{\mu}(\tau,\sigma)$ maps the string into four dimensional
spacetime, and then $g_{\mu\nu}(X)$ can be identified as the background
spacetime metric in which the string is propagating. Note that $\alpha,
\beta$ take values 0, 1 and $\mu,\nu$ do values 0, 1, 2, 3. Here it is 
interesting to note that the effective action of D-string is written in 
a form of the Nambu-Goto action or the Born-Infeld action [12]. 

The classical field equations from the action (1) become
%%%%%%%%%%%%%%%%%%%%%%%%%%%%Equation%%%%%%%%%%%%%%%%%%%%%%%%%%%%%%%%%%%
$$ \eqalign{ \sp(2.0)
0 &= T_{\alpha\beta} = -{2 \over T} {1 \over \sqrt{h}}
     {\delta I \over \delta h^{\alpha\beta}},
\cr
  &= \partial_{\alpha}X^{\mu} \partial_{\beta}X^{\nu}
     g_{\mu\nu}(X) - {1 \over 2} h_{\alpha\beta} h^{\rho\sigma}
     \partial_{\rho}X^{\mu} \partial_{\sigma}X^{\nu}
     g_{\mu\nu}(X),
\cr
\sp(3.0)} \eqno(2)$$
%-----------------------------------------------------------------------
%%%%%%%%%%%%%%%%%%%%%%%%%%%%Equation%%%%%%%%%%%%%%%%%%%%%%%%%%%%%%%%%%%
$$ \eqalign{ \sp(2.0)
0 &= \partial_{\alpha} (\sqrt{h} h^{\alpha\beta} g_{\mu\nu}
     \partial_{\beta}X^{\nu}) - {1 \over 2} \sqrt{h}
     h^{\alpha\beta} \partial_{\alpha}X^{\rho} \partial_{\beta}
     X^{\sigma} \partial_{\mu} g_{\rho\sigma}.
\cr
\sp(3.0)} \eqno(3)$$
%-----------------------------------------------------------------------
Here let us consider the case where the background spacetime metric
$g_{\mu\nu}(X)$ takes a form of the Euclidean Rindler metric
%%%%%%%%%%%%%%%%%%%%%%%%%%%%Equation%%%%%%%%%%%%%%%%%%%%%%%%%%%%%%%%%%%
$$ \eqalign{ \sp(2.0)
ds^2 = g_{\mu\nu}dX^{\mu}dX^{\nu}
     =+g^2z^2dt^2 + dx^2 + dy^2 + dz^2,
\cr
\sp(3.0)} \eqno(4)$$
%-----------------------------------------------------------------------
where $g$ is the redshift factor given by $g = {1 \over 4M}$. 
It is well-known that in the large 
mass limit the Rindler metric can be obtained from the Schwarzschild 
black hole metric in the vicinity of the event horizon and
provides us with a nice playground for examining 
the Hawking radiation in a simple metric form. Here it is important 
to notice that we have performed the Wick rotation with respect to 
the time component since now we would like to discuss the thermodynamic 
properties.

Now one can easily solve Eq.(2) as follows:
%%%%%%%%%%%%%%%%%%%%%%%%%%%%Equation%%%%%%%%%%%%%%%%%%%%%%%%%%%%%%%%%%%
$$ \eqalign{ \sp(2.0)
h_{\alpha\beta} = G (\tau,\sigma) \partial_{\alpha}
                  X^{\mu} \partial_{\beta}X^{\nu} 
                  g_{\mu \nu}(X),
\cr
\sp(3.0)} \eqno(5)$$
%-----------------------------------------------------------------------
where $G(\tau,\sigma)$ denotes the Liouville mode. Next we shall fix 
the gauge symmetries which are the two dimensional diffeomorphisms
and the Weyl rescaling by
%%%%%%%%%%%%%%%%%%%%%%%%%%%%Equation%%%%%%%%%%%%%%%%%%%%%%%%%%%%%%%%%%%
$$ \eqalign{ \sp(2.0)
t(\tau,\sigma) = \tau, \ x(\tau,\sigma) = \sigma, \
                 G(\tau,\sigma) = 1.
\cr
\sp(3.0)} \eqno(6)$$
%-----------------------------------------------------------------------
These gauge conditions express a physical situation where a long string 
stretches along the $x$ direction. Incidentally, in previous works by 
the present author [10,11] we have chosen different gauge conditions 
to realize the stringy paradigm by `tHooft [13].   

At this stage, let us impose a cyclic symmetry
%%%%%%%%%%%%%%%%%%%%%%%%%%%%Equation%%%%%%%%%%%%%%%%%%%%%%%%%%%%%%%%%%%
$$ \eqalign{ \sp(2.0)
y(\tau,\sigma) = y(\tau), \ z(\tau,\sigma) = z(\tau),
\cr
\sp(3.0)} \eqno(7)$$
%-----------------------------------------------------------------------
which is a reasonable assumption from the present physical setting. 
From Eqs.(5), (6) and (7), the world sheet metric takes the form
%%%%%%%%%%%%%%%%%%%%%%%%%%%%Equation%%%%%%%%%%%%%%%%%%%%%%%%%%%%%%%%%%%
$$ \eqalign{ \sp(2.0)
h_{\alpha\beta} = \left(\matrix{g^2 z^2
                  + \dot y^2 + \dot z^2 & 0 \cr
                  0 & 1 \cr} \right),
\cr
\sp(3.0)} \eqno(8)$$
%-----------------------------------------------------------------------
where the dot denotes a derivative with respect to $\tau$.
And the remaining field equations (3) become
%%%%%%%%%%%%%%%%%%%%%%%%%%%%Equation%%%%%%%%%%%%%%%%%%%%%%%%%%%%%%%%%%%
$$ \eqalign{ \sp(2.0)
\partial_{\tau}
({{z^2 } \over \sqrt{h}}) = 0,
\cr
\sp(3.0)} \eqno(9)$$
%-----------------------------------------------------------------------
%%%%%%%%%%%%%%%%%%%%%%%%%%%%Equation%%%%%%%%%%%%%%%%%%%%%%%%%%%%%%%%%%%
$$ \eqalign{ \sp(2.0)
\partial_{\tau}
({{\dot y } \over \sqrt{h}}) = 0,
\cr
\sp(3.0)} \eqno(10)$$
%-----------------------------------------------------------------------
%%%%%%%%%%%%%%%%%%%%%%%%%%%%Equation%%%%%%%%%%%%%%%%%%%%%%%%%%%%%%%%%%%
$$ \eqalign{ \sp(2.0)
\partial_{\tau}
( \ {\dot z \over \sqrt{h}} \ )
- {1 \over \sqrt{h}}g^2 z = 0,
\cr
\sp(3.0)} \eqno(11)$$
%-----------------------------------------------------------------------
where
%%%%%%%%%%%%%%%%%%%%%%%%%%%%Equation%%%%%%%%%%%%%%%%%%%%%%%%%%%%%%%%%%%
$$ \eqalign{ \sp(2.0)
h = g^2 z^2  + \dot y^2 + \dot z^2.
\cr
\sp(3.0)} \eqno(12)$$
%-----------------------------------------------------------------------

Now let us look for the nontrivial solution of the above field equations. 
From Eqs.(9) and (11) we can obtain
%%%%%%%%%%%%%%%%%%%%%%%%%%%%Equation%%%%%%%%%%%%%%%%%%%%%%%%%%%%%%%%%%%
$$ \eqalign{ \sp(2.0)
{z \over \sqrt{h}} = A \cos g (\tau - \tau_0).
\cr
\sp(3.0)} \eqno(13)$$
%-----------------------------------------------------------------------
where $A$ and $\tau_0$ are integration constants. Next Eq.(10) becomes
%%%%%%%%%%%%%%%%%%%%%%%%%%%%Equation%%%%%%%%%%%%%%%%%%%%%%%%%%%%%%%%%%%
$$ \eqalign{ \sp(2.0)
{\dot y \over \sqrt{h}} = B,
\cr
\sp(3.0)} \eqno(14)$$
%-----------------------------------------------------------------------
where $B$ is also an integration constant. The substitution Eqs.(13) and 
(14) into Eq.(12) yields 
%%%%%%%%%%%%%%%%%%%%%%%%%%%%Equation%%%%%%%%%%%%%%%%%%%%%%%%%%%%%%%%%%%
$$ \eqalign{ \sp(2.0)
B = \pm \sqrt{1 - g^2 A^2}.
\cr
\sp(3.0)} \eqno(15)$$
%-----------------------------------------------------------------------
From these equations it is easy to seek for explicit expressions of 
$y(\tau)$ and $z(\tau)$ which are given by
%%%%%%%%%%%%%%%%%%%%%%%%%%%%Equation%%%%%%%%%%%%%%%%%%%%%%%%%%%%%%%%%%%
$$ \eqalign{ \sp(2.0)
y(\tau) =  {C \over g} \tan{g  (\tau - \tau_0)} + y_0,
\cr
\sp(3.0)} \eqno(16)$$
%-----------------------------------------------------------------------
%%%%%%%%%%%%%%%%%%%%%%%%%%%%Equation%%%%%%%%%%%%%%%%%%%%%%%%%%%%%%%%%%%
$$ \eqalign{ \sp(2.0)
z(\tau) = \pm {A C \over B} {1 \over \cos{g  (\tau - \tau_0)}},
\cr
\sp(3.0)} \eqno(17)$$
%-----------------------------------------------------------------------
where $C$ and $y_0$  are the integration constants. 
In order to understand the physical meaning of the solution (17)
more clearly, it is useful to go back to the Lorentzian time $t_L$
%%%%%%%%%%%%%%%%%%%%%%%%%%%%Equation%%%%%%%%%%%%%%%%%%%%%%%%%%%%%%%%%%%
$$ \eqalign{ \sp(2.0)
z(t_L) \ = \pm {A C \over B} {1 \over \cosh{g(t_L - t_{L0})}}.
\cr
\sp(3.0)} \eqno(18)$$
%-----------------------------------------------------------------------
Here note that the Rindler coordinate $(z,t)$ is related to the
Minkowski coodinate $(Z,T)$ by the following transformation:
%%%%%%%%%%%%%%%%%%%%%%%%%%%%Equation%%%%%%%%%%%%%%%%%%%%%%%%%%%%%%%%%%%
$$ \eqalign{ \sp(2.0)
Z = z \cosh gt, \ T = t \sinh gt,
\cr
\sp(3.0)} \eqno(19)$$
%-----------------------------------------------------------------------
thus the above solution (18) corresponds to 
an instanton connecting the past event horizon with the future one with a 
definite constant value $Z = \pm {A C \over B}$.

We now turn to an examination of the thermodynamic properties of the ``world sheet
instanton''(17). It is remarkable that the solution has a periodicity
with respect to the Euclidean time component, $\beta = {2 \pi \over g}$
whose inverse gives us precisely the Hawking temperature $T_H =
{1 \over \beta} = {g \over 2\pi} = {1 \over 8 \pi M}$ of the Rindler
spacetime [9]. This fact implies that the single long string is in 
thermal equilibrium with the heat bath with the Hawking temperature. 

Next let us calculate the entropy carried by the single string 
to the leading order of approximaion by a method developed
by Gibbons and Hawking [14]. By using the relations 
$e^{- \beta F} \sim e^{- I}$ and 
$S = \beta^2 {\partial F \over \partial \beta}$, the above exact 
solutions give us the entropy
%%%%%%%%%%%%%%%%%%%%%%%%%%%%Equation%%%%%%%%%%%%%%%%%%%%%%%%%%%%%%%%%%%
$$ \eqalign{ \sp(2.0)
S =  {1 \over |B|} \ T \ A_H,
\cr
\sp(3.0)} \eqno(20)$$
%-----------------------------------------------------------------------
where $A_H = \int dx dy$ which corresponds to the area of the black
hole horizon and is an infinite quantity as expected since the horizon in 
the Rindler spacetime is infinitely extended plane.  To arrive at the 
Bekenstein-Hawking entropy formula [1,2]
%%%%%%%%%%%%%%%%%%%%%%%%%%%%Equation%%%%%%%%%%%%%%%%%%%%%%%%%%%%%%%%%%%
$$ \eqalign{ \sp(2.0)
S = {1 \over 4G} A_H,
\cr
\sp(3.0)} \eqno(21)$$
%-----------------------------------------------------------------------
we are forced to choose the effective string tension $T$ to be $|B| {1 
\over 4G}$. That is, the string tension should be rescaled by the 
constant factor $|B|$ in comparision to the natural choice ${1 \over 
4G}$. (Of cource there ia an ambiguity in the choice of the constant 
factor ${1 \over 4}$ but this ambiguity is not important now.) What is 
the physical meaning of this factor $|B|$? From the definition Eq.(14) 
it is obvious that this factor is equal to the product of the
absolute value of the velocity of $y$ direction and the inverse of
the redshift factor of the world sheet because of $h = 
h_{00}$. Or as a more suggestive interpretation, 
from Eq.(15) this constant $|B|$ has a close relationship with 
the redshift factor $g$ of the Rindler spacetime. If we select the 
specific value $A^2 = {{1 - g^2} \over g^2}$, we obtain exactly the same 
result $|B| = g$ as E.Halyo et al [7] where it was shown that the 
string tension should be rescaled by this redshift factor $g$
to agree with the black hole entropy. Anyway it is very illuminating that 
the present formalism naturally requires us to take the rescaled string 
tension having a relation with the redshift factor.

It is now well-known that 
the degeneracy of states is given to be $\sim e^{M}$ and $\sim e^{M^2}$ 
for string and black hole respectively. Why have their degeneracies 
equaled in the present formalism? The answer is very simple: Due to the 
redshift factor $g = {1 \over 4M}$ in the Rindler spacetime, the 
degeneracy of a string $\sim e^{M}$ is ``renormalized'' to be that of
the black hole $\sim e^{M^2}$. This fact would be checked explicitly 
if we could succeed in quantizing a string in the Rindler metric and
count the number of degeneracy of states. Also 
the first quantization of a string would lead us to an understanding of 
the renormalization of the string tension. A recent work in this 
direction might help us to consider these problems [15].

To summarize, we have investigated a possibility that a single long 
string can explain the black hole entropy from the viewpoint of string 
theory. This long string has the same Hawking temperature as the 
Schwarzschild (and the Rindler) spacetime and yields the 
Bekenstein-Hawking entropy formula. Even if we have confined ourselves to 
the four dimensional case for simplicity, it is straightforward to 
generalize to arbitrary spacetime dimensions in a similar procedure to the 
ref.[11]. 

We shall make a comment on the future works in the connection of the 
present work. One of the most physically interesting studies is to 
challenge a first quantization of (super)string theory in the Rindler 
spacetime and then see whether the degeneracy of states equals between 
(super)string and black hole. The other is to incorporate the 
antisymmetric tensor and dilaton fields into the present model, i.e. to 
consider the general nonlinear sigma model. It would be amazing if string 
theory not only has the black hole solutions as a part of classical 
solutions but also directly controls their thermodynamical properties 
such as the Hawking radiation and the black hole entropy e.t.c.

\vskip 1cm

\noindent

\leftline{\bf Acknowledgments}

The author would like to thank A.Sugamoto for valuable discussions.
\vskip 12pt
\leftline{\bf References}
\centerline{ } %
\par
\item{[1]} S.W.Hawking,
          Comm. Math. Phys. {\bf 43} (1975) 199. 

\item{[2]} J.D.Bekenstein,
          Nuovo Cim. Lett. {\bf 4} (1972) 737;
          Phys. Rev. {\bf D7} (1973) 2333;
          ibid. {\bf D9} (1974) 3292;
          Physics Today {\bf 33}, no.1 (1980)  24. 

\item{[3]} A.Strominger and C.Vafa, hep-th/9601029. 

\item{[4]} J.Polchinski, Phys. Rev. Lett. {\bf 75} 
          (1995) 4724; J.Polchinski, S.Chaudhuri and 
          C.Johnson, hep-th/9602052.
          
\item{[5]} G.T.Horowitz, gr-qc/9604051 and references 
          therein.

\item{[6]} J.M.Maldacena, hep-th/9605016.

\item{[7]} E.Halyo, A.Rajaraman and L.Susskind, 
           hep-th/9605112.

\item{[8]} A.A.Tseytlin, hep-th/9605091.

\item{[9]} W.G.Unruh, Phys. Rev. {\bf D14} (1976) 870.

\item{[10]} I.Oda, Phys. Lett. {\bf B338} (1994) 165.

\item{[11]} I.Oda, Mod. Phys. Lett. {\bf A10} (1995) 2775.

\item{[12]} R.Leigh, Mod. Phys. Lett. {\bf A4} (1989) 2767; 
            C.Bachas, hep-th/9511043; C.Schmidhuber, 
            hep-th/9601003.
            
\item{[13]}  G.'t Hooft,
          Nucl. Phys. {\bf B335} (1990) 138;
          Physica Scripta {\bf T15} (1987) 143;
          ibid. {\bf T36} (1991) 247.

\item{[14]}  G.Gibbons and S.W.Hawking, Phys. Rev. {\bf D49}
          (1977) 2752; S.W.Hawking, From General Relativity:
          An Einstein Centenary Survey (Cambridge U.P.,
          Cambridge, 1979).
         
\item{[15]}  H.Hata, H.Oda and S.Yahikozawa, hep-th/9512206.
\endpage
%
%=======================================================================%
%
\bye